\documentclass[12pt]{article}
\usepackage{epsfig}
\usepackage{axodraw}
\begin{document}
\renewcommand{\textfraction}{0.0}
\renewcommand{\topfraction}{1.0}
\renewcommand{\bottomfraction}{1.0}

%
\catcode`\@=11
\@addtoreset{equation}{section}
\def\theequation{\thesection.\arabic{equation}}
\catcode`\@=12
\newcommand{\be}{ \begin{equation}}
\newcommand{\ee}{ \end{equation}  }
\newcommand{\bea}{ \begin{eqnarray}}
\newcommand{\eea}{ \end{eqnarray}  }
\newcommand{\bi}{\bibitem}
\newcommand{\rd}{ \mbox{\rm d} }
\newcommand{\rD}{ \mbox{\rm D} }
\newcommand{\re}{ \mbox{\rm e} }
\newcommand{\rO}{ \mbox{\rm O} }
\newcommand{\erf}{\mbox{\rm erf}}
\newcommand{\diag}{\mbox{\rm diag}}

\renewcommand{\floatpagefraction}{0.8}

\def\I{\cite{schro}}
\def\del{\partial}
\def\SF{Schr\"odinger functional }
\def\cb{\bar{c}}
\def\q{\tilde{q}}
\def\c{\tilde{c}}

\def\rmd{{\rm d}}
\def\rmD{{\rm D}}
\def\rme{{\rm e}}
\def\rmO{{\rm O}}
\def\tr{{\rm tr}}

 
\def\gms{g_{\ms}}
\def\gmsbar{g_{\msbar}}
\def\gbar{\bar{g}}
\def\gbarms{\gbar_{\ms}}
\def\gbarmom{\gbar_{\rm mom}}
\def\ms{{\rm MS}}
\def\msbar{{\rm \overline{MS\kern-0.14em}\kern0.14em}}
\def\lat{{\rm lat}}
\def\glat{g_{\lat}}
\def\gbarsf{\bar{g}_{\rm SF}}

\def\alphabar{\alpha}
\def\alphasf{\alpha_{\rm SF}}
\def\alphat{\tilde{\alpha}_0}

 
\def\SU{{\rm SU}(N)}
\def\SUtwo{{\rm SU}(2)}
\def\SUthree{{\rm SU}(3)}
\def\su{{\rm su}(N)}
\def\sutwo{{\rm su}(2)}
\def\suthree{{\rm su}(3)}
\def\pauli#1{\tau^{#1}}
\def\Ad{{\rm Ad}\,}

\begin{titlepage}

\begin{flushright} HUB-EP-98-54 \end{flushright}
\begin{flushright} MPI-PhT/98-70 \end{flushright}

\vspace{1.cm}
\begin{center}
{\LARGE Two-Loop Computation of the Schr\"odinger Functional
in Pure SU(3) Lattice Gauge Theory\\}
\vskip 1 cm
\vbox{
\centerline{
\epsfxsize=2.5 true cm
\epsfbox{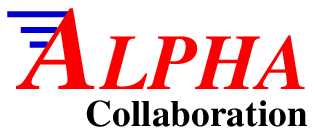}}
}
\vskip 1 cm

%
{\large Achim Bode and Ulli Wolff\\
Institut f\"ur Physik, Humboldt Universit\"at\\
Invalidenstr. 110, D-10099 Berlin, Germany\\}
\vspace{.5cm}
{\large Peter Weisz\\
Max-Planck-Institut f\"ur Physik\\
F\"ohringerring 6, D-80805 M\"unchen, Germany}
\end{center}
\vspace{.5cm}
\thispagestyle{empty}
\begin{abstract}\normalsize
We compute  the Schr\"odinger functional (SF) 
for the case of pure SU(3) gauge theory 
at two-loop order
in lattice perturbation theory.
This allows us to extract the three-loop $\beta$-function in the SF-scheme.
These results are required to compute the running coupling,
the $\Lambda$-parameter
and quark masses by finite size techniques with negligible
{\it systematic} errors.
In addition, we can now implement two-loop O($a$) improvement
in simulations and extend and study series
in alternative (``tadpole-improved'') bare couplings.
\end{abstract}

\end{titlepage}

\section{Introduction}

Traditionally, in computer simulations of lattice QCD 
physicists have been concerned
with the computation of low energy quantities
which are
inaccessible to perturbation theory.
Typical examples are
ratios of hadron masses, the static quark potential at large separations
and glueball masses. At high
energy, where QCD was first established as a promising theory
of strong interactions,
perturbation theory in the dimensionally regularized
continuum theory is the standard computational tool. 
Here it provides predictive power by reproducing a host of
high energy data in terms of a renormalized running coupling
and running quark masses at that energy scale.
As a result
the coupling in the $\msbar$-scheme at the scale of the mass of
the $Z$-boson is fixed to be
$\alpha_{\msbar}(M_Z) \approx 0.11$.
For a long time the two energy sectors --- though they are believed to
belong to one and the same
theory --- have coexisted rather independently. Only in recent years
was the task of connecting them by {\em computing} $\alpha_{\msbar}$ in
the theory that is completely parameterized at low energy, taken up
(see \cite{rev1,rev1a,rev2} for reviews).

Referring also to low energy this result has to be obtained
by simulation on the lattice.
The extraordinary difficulty of this computation is caused
by the fact that we seem to have to simultaneously
control the continuum limits of
two widely separated physical energies. 
Naively this requires unfeasibly large lattices.
The problem is circumvented by the somewhat indirect strategy
developed by the ALPHA-collaboration in recent years \cite{rev1a,rev2}.
Here we numerically trace the evolution of 
the coupling $\alphasf$ defined by the \SF from
low to high energy. 
While not being a directly physical
quantity, it is nevertheless expected to
possess a well defined continuum limit
and the motivation for this particular choice 
is explained in \cite{LSWW}.
Due to its definition on a finite system, it runs with the size $L$.
Once its evolution is known, it remains to connect 
both its low and high energy values
to more experimentally accessible quantities. 
At low energy (i.e. at large physical volume) this is done
by numerical simulation, which however no longer involves
unmanageable scale ratios. At high energy, $\alphasf$ on small
systems can be perturbatively related to $\alpha_{\msbar}$.
It has turned out that with the precision that is both experimentally
required and achievable in the various simulation steps, this connection
is needed at {\em two-loop} order.
Due to the difficult structure of lattice perturbation theory
one usually has to content oneself with one-loop results.
In fact, the results presented here and its analogue for gauge
group SU(2) \cite{NW}
are the only lattice two-loop results of
renormalized quantities known to us.

The perturbative relation between $\alphasf$ and $\alpha_{\msbar}$
is established in two steps through the bare lattice coupling $g_0$ as
an intermediate quantity. The two-loop relation between $\alpha_{\msbar}$
and $g_0$ has been worked out in \cite{LWms} for Yang-Mills theory with
gauge group SU($N$) for general $N$. The present paper completes the
two loop relation for quenched QCD by deriving the connection
between $g_0$ and $\alphasf$ for the realistic case of SU(3)\footnote{
Results reported here are based on ref. \cite{phd} where more details
can be found.}.
The analogous SU(2) calculation was reported in \cite{NW}.
We note that here Feynman diagrams have to be evaluated and
extrapolated numerically. This cannot be done for general $N$, and
the more complex group structure of SU(3) represents a very significant
complication.

After this introduction, we define the problem for SU(3) and 
give a few hints to details of the computation
in section 2. In section 3 we compile results for finite
lattices and analyze and extrapolate them. Finally in section 4 some
applications of our results are reviewed. Some preliminary results
of this work were already given in \cite{achimlat97}.

\section{Perturbation theory for gauge group SU(3)}

In the \SF approach \cite{LNWW} we compute the free energy $\Gamma$ given
by the lattice path integral
\be
 e^{-\Gamma} = \int D[U] e^{-S(U)}.
\label{gammadef}
\ee
The geometry is a finite box of size $L$
with periodic boundary conditions in the
spatial directions and Dirichlet boundary conditions in
time.  In the lattice realization
the latter means that the spatial links $U(x,k), k=1,2,3$,
are frozen to fixed SU(3)-elements
for the layers at time coordinate $x^0=0,L$,
\bea
U(x,k)|_{x^0=0}&=& \exp(a C)=
\exp\Big[\,\frac{ia}{L}\,\diag(\phi_1,\phi_2,\phi_3)\Big],
\label{surfaceB}\\
U(x,k)|_{x^0=L}&=& \exp(a C')=
\exp\Big[\,\frac{ia}{L}\,\diag(\phi'_1,\phi'_2,\phi'_3)\Big].
\label{surfaceT}
\eea
Here $a$ is the lattice spacing, and
$\phi_i, \phi_i'$ are certain \cite{LSWW} 
dimensionless numbers that depend on
one free parameter $\eta$ that we shall vary later on
to probe the system.
Both our perturbative calculation here
and simulations reported in \cite{LSWW} and  more recently in
\cite{newsims,CLSW} are restricted
to the choice ``A'' of \cite{LSWW} for these numbers.
All other links interior to the box are integrated over with the
invariant SU(3) measure in (\ref{gammadef}).
The action is defined by the usual sum over oriented plaquettes,
\be
S(U) = \frac1{g_0^2} \sum_p w(p) \, \tr(1-U_p).
\ee
The weight $w(p)$ is unity for all plaquettes
except those at the boundary that
contain the time-direction and one of
the frozen spatial links where we put
\be \label{impweight}
w(p) = c_t(g_0) = 1 + c_t^{(1)}g_0^2 + c_t^{(2)}g_0^4 + \ldots
\ee
The freedom of adjusting this weight is
required for improvement of $\rmO(a)$ lattice artefacts
that are otherwise introduced by the surfaces.

We set $\alphasf = \gbar^2/4\pi$ and
the SF-coupling $\gbar$ is defined from the response in the energy $\Gamma$
to infinitesimal changes in the surface fields by varying $\eta$,
\be
\gbar^2 = k/\Gamma' \; ,
\ee
where $\Gamma'$ is the derivative with respect to $\eta$ at $\eta=0$
and $k$ is a constant.
It is fixed by normalizing the leading term in the perturbative expansion
\be \label{gbarpert}
 \gbar^2(L) = g^2_0 + m_1(L/a) g^4_0 + m_2(L/a) g^6_0 + \ldots
\ee
The one loop coefficient $m_1$ was computed in ref.~\cite{LSWW}.   
Our objective here is the computation of $m_2$ for gauge group SU(3).
In order to determine the improvement coefficients in (\ref{impweight})
we need to explicitly exhibit the dependence of the coefficients
$m_i$ on them. 
Inspection of the structure of the contributions
leads us to write
\bea
  m_1 &=& m_1^a + c_t^{(1)} m_1^b, \\
  m_2 - m_1^2 &=& m_2^a + c_t^{(1)} m_2^b 
  + \left[c_t^{(1)}\right]^2 m_2^c + c_t^{(2)} m_2^d.
  \label{impabcd}
\eea

To work out the perturbative expansion of $\Gamma$ 
we fix
the gauge as discussed in \cite{LNWW}. 
It is an important advantage of the SF-framework that there is
a unique background field of minimal action that interpolates
between the surface values (\ref{surfaceB}), (\ref{surfaceT}),
\be
V(x,k)=\exp[a (C + (C'-C)x^0/L)], \; V(x,0)= 1,
\label{bgfield}
\ee
and all fluctuation modes receive gaussian damping.
To deal with the SU(3) Lie-algebraic structure of the 
fluctuations it is advantageous to introduce a suitable basis.
We used generators proportional to
\bea
T^0 &=& \diag(0,1,-1) \propto (C'-C), \nonumber\\
T^1 &=& \diag(2,-1,-1),\\
(T^{\alpha\beta})_{kl} &=& \delta^{\alpha}_k  \delta^{\beta}_l;
\quad (\alpha\beta)=(12),(13),(23),
\nonumber
\eea
such that $T^0,T^1$ span the Cartan subalgebra. The $T^{\alpha\beta}$
and their adjoints
are natural generalizations of the ladder-operators in SU(2).
Now covariant derivatives in the background field acting in the adjoint
representation, 
--- an ubiquitous
structure in our perturbation theory --- are diagonalized in group space,
\be
[T^0,T^{\alpha\beta}] = h_{\alpha\beta} T^{\alpha\beta}
\ee
with numbers $h_{\alpha\beta}$.
In this basis the numerical construction of the propagators 
in the background field
is reduced to the SU(2) case. We just have to apply the procedure
described in section 5 of ref. \cite{NW} several times for various effective
background field parameters.

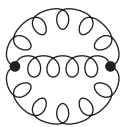
\begin{figure}[ht]
\begin{center}
\begin{picture}(170,40)(0,17)
\SetOffset( 50,10)
\GlueArc(30,20)(18,0,180){4}{5}
\GlueArc(30,20)(18,180,360){4}{5}
\Vertex (12,20){2}
\Vertex (48,20){2}
\Gluon  (12,20)(48,20){4}{4}
\end{picture} 
\caption[]{\label{bigmacfigure} The most computer-time intensive diagram, the
big-mac with three gluon propagators.}
\end{center}
\end{figure}

While the vertices for SU(2) were still `hand-coded' in Fortran,
the SU(3) structure was so much more complicated that we let
Maple do the work. The three-gluon vertex, for instance, consists of
532 terms. 
As for SU(2), we performed the summations
in position space, so that the most demanding big-mac diagram
shown in fig.~\ref{bigmacfigure} has only O($(L/a)^5$) instead of 
O($(L/a)^8$) terms in
momentum space. 
Its total number of contributions for $L/a=32$ 
after all possible reductions is however still
about $5 \cdot 10^{12}$. 
Along with each vertex amplitude and propagator we computed
their exact $\eta$-derivatives to form $\Gamma'$.
To enhance the confidence in our numbers
we took the same precautions as in \cite{NW}: test of all symmetries
before using them to reduce the number of terms by a factor of 36 altogether, 
check of gauge independence and
comparison with a slow but independent code for the  smallest lattices.

\section{Extraction of results}

Some of the coefficients in (\ref{impabcd}) can be given in closed form.
Since the background field (\ref{bgfield}) leads to constant plaquettes
it is easy to see that
\be
m_1^b = m_2^d = -\frac{2a}{L}.
\ee
The tree-level contribution $m_2^c$ can be evaluated
in closed form.
The ensuing expression is however a complicated 
function of the background field parameters
and for our purposes
the leading behavior is sufficient
\be
m_2^c = \frac{2a}{L} -\frac{4a^2}{L^2}+ \rmO(a^5).
\ee
The remaining contributions $m_1^a$, $m_2^a$ and $m_2^b$ are tabulated
in table~\ref{mtable}.
\begin{table}[ht]
\begin{center}
\begin{tabular}{|c|l|l|l|}\hline
\rule[-0.5ex]{0ex}{2.6ex}
 $L/a$ &\multicolumn{1}{|c|}{$m_1^a$}
       &\multicolumn{1}{|c|}{$m_2^a$}
       &\multicolumn{1}{|c|}{$m_2^b$}         \\ \hline
  4 &0.529761318876020&0.052943274533542&0.106270501583520\\ 
  5 &0.564680831258396&0.056838593961882&0.100438272480780\\ 
  6 &0.593212325999814&0.05975069070789	&0.09439258604949 \\ 
  7 &0.617378117870096&0.06197806027570	&0.08867849757097 \\ 
  8 &0.638244418293146&0.06374136571959	&0.08343000201173 \\ 
  9 &0.656541534280898&0.06518532139185	&0.07867547537704 \\ 
 10 &0.672800386347018&0.06640139951338	&0.07439325408456 \\ 
 11 &0.687413016523170&0.06744855730398	&0.07054052556373 \\ 
 12 &0.700673174318993&0.06836628672255	&0.06706893587917 \\ 
 13 &0.712804791119344&0.0691821005300	&0.06393184996835 \\ 
 14 &0.723981656094021&0.0699158333750	&0.06108717717361 \\ 
 15 &0.734340962964557&0.0705822007834	&0.05849809311936 \\ 
 16 &0.743992719364831&0.0711923868147	&0.0561328521011  \\ 
 17 &0.753026381769789&0.0717550680760	&0.0539642582909  \\ 
 18 &0.761515615421704&0.0722770966172	&0.0519690528035  \\ 
 19 &0.769521768768977&0.0727639684486	&0.0501273255443  \\ 
 20 &0.777096451745942&0.0732201529077	&0.0484219921318  \\ 
 21 &0.784283478608195&0.0736493291767	&0.0468383453838  \\ 
 22 &0.791120352757074&0.0740545593455	&0.0453636778366  \\ 
 23 &0.797639416324361&0.0744384171842	&0.0439869669949  \\ 
 24 &0.803868750843028&0.0748030854189	&0.0426986139530  \\ 
 25 &0.809832890636312&0.0751504302247	&0.0414902265431  \\ 
 26 &0.815553393561204&0.075482058988	&0.0403544392531  \\ 
 27 &0.821049301867167&0.075799365607	&0.0392847633581  \\ 
 28 &0.826337517515302&0.076103566396	&0.0382754618367  \\ 
 29 &0.831433110259494&0.076395728812	&0.0373214446266  \\ 
 30 &0.836349572396415&0.076676794651	&0.0364181805963  \\ 
 31 &0.841099030858248&0.076947598928	&0.0355616232918  \\ 
 32 &0.845692424917554&0.077208885368	&0.0347481480648  \\ \hline
\end{tabular}
\end{center}
\caption[]{\label{mtable}List of $L/a$-dependent coefficients  
$m_1^a$, $m_2^a$ and $m_2^b$.
Numerical errors are beyond the quoted digits.}
\end{table}
The generically expected behavior for 
2-loop lattice Feynman diagrams suggests an asymptotic expansion
\be
m_2^a = \sum_{n=0}^{\infty} [r_n+s_n \ln(L/a) + t_n \ln^2(L/a)](a/L)^n.
\ee
For $m_1^a$ we expect the same structure without $\ln^2$-terms and the same
is true for the two-loop diagrams contributing to $m_2^b$ 
which in addition starts at $n=1$ only.
The coefficients of these expansions are extracted from the series 
of finite lattices
and errors are estimated by the blocking procedure described in \cite{LWblock}.
The Callan-Symanzik equation
\be
L \frac{\del}{\del L} \gbar(L) = - \beta(\gbar) = b_0 \gbar^3 + b_1 \gbar^5
+ b_2 \gbar^7 + \ldots
\label{SFbetafunction}
\ee
implies $t_0=0$ and fixes all logarithmic divergences in terms of the
universal coefficients
\be
b_0=\frac{11}{(4\pi)^2}, \; b_1=\frac{102}{(4\pi)^4}. 
\ee
These values were first verified to high
accuracy from the data and then assumed to be exact to extract subleading
terms. Cancellation of all terms of O($a$)  
in $m_1$ and $m_2$ is achieved to our numerical accuracy by choosing
\be
c_t = 1 - 0.08896(23) g_0^2 - 0.0301(25) g_0^4 + \rmO(g_0^6).
\label{ctimp}
\ee
With these values, referred to as 2-loop perturbative O($a$) improvement,
we finally quote 
\bea
m_1 &=& 2 b_0 \ln(L/a) + 0.36828215(13) + \rmO(a^2) , \\
m_2-m_1^2 &=& 2 b_1 \ln(L/a) + 0.048085(63) + \rmO(a^2) ,
\eea
for the relation (\ref{gbarpert}) between $g_0^2$ and $\gbar^2$.

The two-loop connection between two different
couplings determines the difference
between the non-universal three-loop coefficients of their 
respective $\beta$-functions.
Hence we can employ the results in \cite{LWms}, where use is made of
the three-loop $\beta$-function in the $\overline{\rm MS}$-scheme to determine
\be
b_2 = \frac{0.4828(88)}{(4\pi)^3}
\ee
for the third coefficient in (\ref{SFbetafunction}).
This is of the order of magnitude of the effective two-loop
coefficient $b_2^{\rm eff}=1.5(8)/(4\pi)^3$ which was used in 
ref.~\cite{LSWW} to represent the data.

A central quantity in the ALPHA-collaboration's approach is the step scaling
function which generalizes the $\beta$-function to finite rescalings,
\be
  \sigma(s,u) = \gbar^2(s L) \left. \right|_{u=\gbar^2(L)}.
\ee
On the lattice, it emerges as the continuum limit of a
finite lattice spacing approximant $\Sigma$,
\be
  \sigma(s,u) = \lim_{a\to 0} \Sigma(s,u,a/L).
\ee
All perturbative information about the convergence speed for $s=2$ is
conveniently summarized in coefficients $\delta_1^{(k)}$ and 
$\delta_2^{(k)}$ defined by
\be
  \delta^{(k)} (u,a/L) = {\Sigma(2,u,a/L)-\sigma(2,u) \over \sigma(2,u)}=
  \delta_1^{(k)} (a/L) \, u + \delta_2^{(k)} (a/L) \, u^2 + \ldots ,
  \label{deltas}
\ee
where the upper index $k=0,1,2$ refers to varying the degree of improvement
by using an action with
one, two or three terms in the expansion 
of $c_t$ in (\ref{ctimp}). Our results are collected in table~\ref{deltatable}.
\begin{table}[ht]
\begin{center}
\begin{tabular}{|c|ccccc|}\hline
\rule[-0.5ex]{0ex}{3.2ex}
 $L/a$
&$\delta_1^{(0)}$
&$\delta_1^{(1)}$
&$\delta_2^{(0)}$
&$\delta_2^{(1)}$
&$\delta_2^{(2)}$\\\hline
$  4  $&$ 0.01192 $&$ -0.01023  $&$ 0.00642  $&$ 0.00577 $&$ -0.00168 $\\   
$  5  $&$ 0.01155 $&$ -0.00617  $&$ 0.00514  $&$ 0.00501 $&$ -0.00095 $\\   
$  6  $&$ 0.01089 $&$ -0.00387  $&$ 0.00412  $&$ 0.00435 $&$ -0.00061 $\\   
$  7  $&$ 0.01004 $&$ -0.00262  $&$ 0.00334  $&$ 0.00382 $&$ -0.00043 $\\   
$  8  $&$ 0.00918 $&$ -0.00189  $&$ 0.00275  $&$ 0.00341 $&$ -0.00032 $\\   
$  9  $&$ 0.00841 $&$ -0.00144  $&$ 0.00230  $&$ 0.00307 $&$ -0.00024 $\\   
$ 10  $&$ 0.00773 $&$ -0.00113  $&$ 0.00195  $&$ 0.00279 $&$ -0.00019 $\\   
$ 11  $&$ 0.00714 $&$ -0.00091  $&$ 0.00168  $&$ 0.00256 $&$ -0.00015 $\\   
$ 12  $&$ 0.00663 $&$ -0.00075  $&$ 0.00145  $&$ 0.00236 $&$ -0.00012 $\\   
$ 13  $&$ 0.00618 $&$ -0.00063  $&$ 0.00126  $&$ 0.00219 $&$ -0.00010 $\\   
$ 14  $&$ 0.00579 $&$ -0.00054  $&$ 0.00111  $&$ 0.00205 $&$ -0.00008 $\\   
$ 15  $&$ 0.00544 $&$ -0.00046  $&$ 0.00098  $&$ 0.00192 $&$ -0.00007 $\\   
$ 16  $&$ 0.00513 $&$ -0.00040  $&$ 0.00087  $&$ 0.00180 $&$ -0.00006 $\\\hline
\end{tabular}
\caption[]{\label{deltatable}The perturbative lattice effects.}
\end{center}
\end{table}

\section{Applications}

The perturbative coefficients, whose computation we described in the previous
sections, have already found some applications which we now briefly discuss.
Once the SF-coupling $\gbar$ is known for small box-size, it is to be converted
to the $\overline{\rm MS}$-coupling at high energy. One could conventionally
choose the mass of the neutral weak boson $M_Z$ as a scale here.
Another choice --- attractive for asymptotically free theories ---
is to extract the $\Lambda$-parameter which is simply related to the behavior
at asymptotically large energy.
It is a renormalization group invariant given by
\be
\Lambda_{\rm SF} = L^{-1} (b_0 \gbar^2)^{-b_1/(2b_0^2)} \rme^{-1/(2b_0\gbar^2)}
\times \exp\left\{ -\int_0^{\gbar} \rmd g \left[ \frac1{\beta(g)}
+\frac1{b_0 g^3} -\frac{b_1}{b_0^2 g}
\right]\right\}.
\label{LambdaSF}
\ee
The conversion to $\Lambda_{\msbar}$ then amounts to an 
additional known factor. 
If we insert a small $\gbar(L)$ belonging to a very small $L$ into formula (\ref{LambdaSF}),
then the exponentiated integral is close to unity. In \cite{newsims} the $\Lambda$-parameter in the $\msbar$
scheme at zero flavor number has been estimated to be
\be
\Lambda_{\msbar}^{(0)} = 251(21) {\rm MeV}.
\ee
The quoted error is the statistical error from the simulations of the couplings.
The three-loop term of the $\beta$-function has been used and this renders
the systematic error due to further orders negligible.
Without the three loop term, a hard to quantify systematic error of the order
of the statistical one would have remained.

In \cite{GSW} the continuum limit of $L_{\rm max}/r_0 = 0.718(16)$ is
computed.  
The box-size $L_{\rm max}$ is defined \cite{LSWW} via the
\SF by $\gbar^2(L_{\rm max}) = 3.48$.  
The distance $r_0$ is fixed
though the relation $r_0^2 F(r_0)=1.65$, where $F$ is the force
between static quarks \cite{rnot}. 
It is expected to be close 0.5 fm in physical units.
Some of the lattice artefacts derive from the definition of $L_{\rm max}$
and are expected to be O($a$). Series of simulations have been conducted
both with the one-loop and two-loop forms of $c_t$, eq.~(\ref{ctimp}).
In fig.~5 of \cite{GSW} the effect of $c_t^{(2)}$ is very clearly visible.
It practically eliminates the O($a$) term in the continuum extrapolation
which thus becomes more reliable and accurate.

We conclude with some remarks on the numerical quality of perturbative expansions
in the bare coupling, $\alpha_0=g_0/4\pi$, or  improved bare couplings \cite{tadpole}
like $\alpha_P = \alpha_0/P$, where $P$ is the average plaquette (in infinite volume)
for the corresponding bare coupling. There are few occasions to investigate
such series beyond one-loop order as we can do now in the case of $\alphasf$.
The well-known problem with bare perturbation theory shows up in the
size of the coefficients at scale $q=1/a$,
\be
\alphasf(1/a) = \alpha_0 + 4.628 \,\alpha_0^2 + 29.0 \,\alpha_0^3 + \rmO(\alpha_0^4).
\ee
If we choose a scale $s/a$ where the 1-loop term is cancelled,
then $s$ is unnaturally large and the next term is still not small
\be
\alphasf(14.06/a) = \alpha_0 + 4.18 \,\alpha_0^3 + \rmO(\alpha_0^4).
\ee
If we change to the boosted coupling the analogous formulas look somewhat better,
\bea
\alphasf(1/a) &=& \alpha_P + 0.439 \,\alpha_P^2 + 
2.43 \,\alpha_P^3 + \rmO(\alpha_P^4),\\
\alphasf(1.285/a) &=& \alpha_P + 1.91 \,\alpha_P^3 + \rmO(\alpha_P^4).
\eea
The same tendency is obtained if we minimise the two-loop coefficient
instead of cancelling the one-loop term
because the scale ratio $r$ between the former and latter method  
comes out to be
\be
r=\exp(b_1/(4b_0^2)) 
\ee
which is close to unity.
With typical couplings like $\alpha_P \simeq 0.11$ for $\beta = 6.5$ it still is
difficult to be confident about systematic errors incurred by
truncating the series.

As the ALPHA collaboration moves to the inclusion of two massless flavors of quarks
into simulations of $\alphasf$ we plan to extend the present two-loop calculation by the
required diagrams with quarks. We hope to report on this in the near future.

\end{document}